# Enforcing Transparent Access to Private Content in Social Networks by Means of Automatic Sanitization

Alexandre Viejo, David Sánchez[1]

*UNESCO Chair in Data Privacy, Department of Computer Science and Mathematics,*
*Universitat Rovira i Virgili, Avda. Països Catalans, 26, 43007 Tarragona, Spain*
*{alexandre.viejo, david.sanchez}@urv.cat*

**Abstract**

Social networks have become an essential meeting point for millions of individuals willing to publish and consume huge quantities of heterogeneous information. Some studies have shown that the data published in these platforms may contain sensitive personal information and that external entities can gather and exploit this knowledge for their own benefit. Even though some methods to preserve the privacy of social networks users have been proposed, they generally apply rigid access control measures to the protected content and, even worse, they do not enable the users to understand which contents are sensitive. Last but not least, most of them require the collaboration of social network operators or they fail to provide a practical solution capable of working with well-known and already deployed social platforms. In this paper, we propose a new scheme that addresses all these issues. The new system is envisaged as an independent piece of software that does not depend on the social network in use and that can be transparently applied to most existing ones. According to a set of privacy requirements intuitively defined by the users of a social network, the proposed scheme is able to: (i) automatically detect sensitive data in users' publications; (ii) construct sanitized versions of such data; and (iii) provide privacy-preserving transparent access to sensitive contents by disclosing more or less information to readers according to their credentials toward the owner of the publications. We also study the applicability of the proposed system in general and illustrate its behavior in two case studies.

---

[1] Corresponding author. Address: Departament d'Enginyeria Informàtica i Matemàtiques. Universitat Rovira i Virgili. Avda. Països Catalans, 26. 43007. Tarragona. Spain. Tel.: +34 977 559657; E-mail: david.sanchez@urv.cat

# 1. Introduction

A social network is a virtual environment powered by Web 2.0 technologies that enables users to publish and share all kinds of information and services with a global audience. Well-known platforms such as Facebook or Twitter have brought together more than 800 and 100 million active users respectively that *generate* and *consume* social contents (McMillan, 2011).

The Consumer Reports'2010 State of the Net analysis (Consumer Reports National Research Center, 2010) stated that more than half of the users of social networks share private information about themselves online. This shows that the published content may contain sensitive personal data and, thus, it may represent a serious privacy threat (Velásquez, 2013). For example, some entities may exploit that knowledge to obtain benefits for their business (e.g. personalized spamming, phishing, etc.) (Zhang, Sun, Zhu, & Fang, 2010); recruiters may use it to hire or discard candidates (Snowdon, 2011); or it can even be used by regular people to perform bullying in the workplace or in the classroom (D'Arcy, 2011).

In recent years, social network users have been increasingly aware of that situation and studies have shown that their privacy concerns have negatively affected the way they use these applications (Staddon, Huffaker, Larking, & Sedley, 2012). Specifically, it has been reported that privacy-aware users spend less time posting and/or commenting social content. Due to the fact that an important part of the business model of the social network operators depends on the use of the social content generated by the users to attract advertisers (Crimes, 2012), the privacy concerns of the privacy-aware users represent a significant problem to the economic success of these platforms.

In order to minimize this issue, some social network operators have implemented limited privacy settings that allow users to decide who will have access to certain contents such as their profile attributes or their published messages. For example, Twitter allows their users to keep their messages (i.e., tweets) public (this is the default setting) or to protect them. Public tweets are visible to anyone, whether or not they have a Twitter account, while protected tweets are only visible to approved Twitter followers. On the other hand, Facebook enables the owners of the social data to provide access to different groups of users (anyone, friends, friends of friends, co-workers, etc.) or even to single users.

Even though these privacy-preserving mechanisms represent a certain improvement with respect to the user's privacy, they have been criticized in the literature due to the following three essential problems: (i) these privacy settings are generally not sufficiently understood by the average users who seldom change the default configuration that generally makes most of the user information public (Bilton, 2010; Stern & Kumar, 2014; Van Eecke & Truyens, 2010); (ii) users are not informed about the privacy risks that their published data may cause and, therefore, they may find difficulties in defining effective privacy settings over their data (Wang, Nepali, & Nikolai, 2014); and (iii) these privacy settings, in any case, do not prevent social network operators from gathering sensitive user data and exploiting it to obtain economic benefits from advertisers or other entities (Crimes, 2012; Viejo, Castellà-Roca, & Rufián, 2013; Wilson, 2011).

Those problems have been already identified in the literature. Works such as (Becker & Chen, 2009; Talukder, Ouzzani, Elmagarmid, Elmeleegy, & Yakout, 2010; Wang, et al., 2014) have provided solutions that measure the users' privacy exposure by means of their published profile attributes (e.g., address, political views, religious views, etc.) and warn them when their exposure level is too high; also, schemes such as (Viejo, et al., 2013) or (Conti, Hasani, & Crispo, 2011) have been designed to protect the sensitive user data from unauthorized entities, which include social network operators. Nevertheless, despite the efforts of the scientific community, there are two related issues that still have room for improvement: (i) there is not an effective mechanism that enables users to measure the degree of sensitivity of their textual publications (i.e., how dangerous is a certain tweet/timeline post from the privacy point of view?); and (ii) the privacy settings usually implemented in social networks offer a very *rigid* access control, that is, these methods manage the protected data as an indivisible element and, as a result, users either get full access to the whole protected data or they cannot obtain anything.

According to these points, new privacy-preserving mechanisms for social networks should be designed and, ideally, these new schemes should automatically warn the users about the privacy risks inherent to their textual publications and propose ways to reduce those hazards. Methods that offer access to contents should also be more flexible and transparent to the users. Finally, those systems should be deployed and managed by the users themselves to prevent the social network operators from gathering social data at will.

## 1.1. Previous work

As discussed earlier, the privacy settings implemented by social networks operators have been widely disqualified by the scientific community and some alternative approaches have been presented to protect the privacy of the users in a more effective way. We next review the most relevant ones.

Developing user privacy policies (as a contract that specifies who can access to a certain resource) for social networks and enforcing their application to ensure the proper protection of private data is a well-known approach used in works such as (Aimeur, 2010; C. Carminati, Ferrari, Heatherly, Kantarcioglu, & Thuraisingham, 2011; Cheek & Shehab, 2012; Dhia, Abdessalem, & Sozio, 2012). A main shortcoming of this approach is that it requires the social network operators to implement such policies. Due to the fact that it is not clear which benefits would get an operator that implements those methods, it can be assumed that, for the moment, this approach is unlikely to be applied. Moreover, in any case, this approach does not prevent the social network operator from obtaining the protected data.

A more straightforward way to achieve this would be to create new privacy-enabled social networks specifically designed to apply strong privacy policies and to really preserve the privacy of their users. However, as stressed in (Baden, Bender, Spring, & Bhattacharjee, 2009), it is unrealistic to assume that a new brand social network can replace the well-known existing ones. In order to deal with this situation, the authors of this last paper present *Persona,* a social network integrated into Facebook as an application to which users log in through a Firefox extension. The users of Persona define a privacy policy that manages the access to their personal information. As a result, only the users with the appropriate access rights can get the protected data. Nevertheless, this tool is only a Facebook application that can be easily removed by the social network operator from the applications directory.

In spite of the low success probabilities of brand new social networks, there are some proposals in the literature that focus on designing new privacy-enabled social networks based on completely distributed architectures. Diaspora[2] is probably the clearest example of this approach; however, other systems such as (Cutillo, Molva, & Strufe, 2009; Nilizadeh, Jahid, Mittal, Borisov, & Kapadia, 2012; Vu, Aberer, Buchegger, & Datta, 2009) can also be found in the literature. This kind of schemes allows users to install and manage their own personal web server to store all their data (e.g., photos, videos, etc.). Since each user controls her data server, she retains full ownership over the shared content and is not subjected to changing privacy policies and sell-outs to third parties (Vaughan-Nichols, 2010). The general idea of this approach is interesting but, as explained above, it is quite unlikely that those distributed social networks will attract enough users to become a real alternative.

The last approach considered in the literature is based on replacing the data to be protected by fake information that looks realistic in front of the social network operator. Under this approach, fake information is published in the social network while the real data is protected and stored in a way that the social network operator cannot obtain it. All the schemes in this category provide a software component that is in charge of transparently showing the real information when an authorized user browses the protected profile of another user. Note that the use of fake data that looks realistic is essential in those schemes due to the fact that the terms of service of well-known social networks usually state that it is forbidden to completely obfuscate the published personal data. In fact, Facebook has banned users who have violated those terms (Scoble, 2008). This behavior prevents straightforward solutions such as using cryptography to cipher text or attributes before publishing them.

---

[2] http://joindiaspora.com (last accessed: January, 2016).

Some researchers have followed this approach (Conti, et al., 2011; Luo, Xie, & Hengartner, 2009; Viejo, et al., 2013) . The main difference between these schemes is the place where the real data is stored. In (Luo, et al., 2009), the real data is stored in an external centralized infrastructure that must be honest and always available. In (Conti, et al., 2011), the authors state that assuming the existence of a fully-honest and always-available third party server is unrealistic; therefore, the authors propose to locally store the real data in the computers of authorized friends. This behavior presents two main problems: (i) users are required to always connect to the social network by using the computer that locally stores the real data; and (ii) whenever a user modifies her protected information, it has to be individually sent to all the authorized friends for updating. Given that the society has already entered the age of ubiquitous connectivity, the first issue is a very strong drawback. Regarding the second issue, it is quite inefficient in terms of bandwidth usage and it also might generate an unstable situation where not all the authorized recipients would have access to the up-to-date information. Moreover, users might not be comfortable with storing personal data from other users. In order to solve the availability issues of (Luo, et al., 2009) and the problems related to the use of the distributed solution presented in (Conti, et al., 2011), the authors of (Viejo, et al., 2013) store both fake and protected real information in the servers of the social network. Nevertheless, the protected information is properly hidden in images using *steganography* (i.e., the practice of concealing a message, image, or file within another message, image, or file). These images are accessible by anyone browsing the social network accounts of the users, but: (i) they have to be aware that a certain image contains hidden data; and (ii) the data embedded in the image is properly encrypted, therefore, only the individuals with the right cryptographic keys are capable of obtaining the real data.

In any case, all these analyzed proposals require the users to: (i) understand which publications are risky and should be protected; and (ii) manually select who can access to a certain protected resource. Both requirements clearly threaten the applicability of the solutions proposed so far. Note that if a user is not aware of the sensitivity of a certain piece of data, she may publish it publicly on purpose without using any privacy-preserving tool. Therefore, providing a privacy-preserving solution that quantifies the privacy risks the contents to be published and provides a transparent access to them would represent and important advance.

Another limitation of current works is the rigid access to protected elements. The analyzed schemes enable allowed users to fully access to the protected elements while any not allowed user will not be able to obtain any data. Even though this behavior is functional, we stress that providing more flexible techniques can improve the usefulness of the published data without jeopardizing the privacy of the users. For example, a more flexible proposal may provide a certain degree of information (i.e., from more specific to more general) depending on the credentials of the user. In this way, a user with limited credentials could still get a certain amount of information (clearly, more restricted than what can be obtained by a user with full credentials) instead of getting no information at all.

### 1.2. Contributions and plan of this paper

In this paper, we address the issues discussed above by proposing a new privacy-preserving scheme that enables the users of already deployed social networks to transparently preserve their privacy in front of other users, third parties and even the social network operator itself.

The proposed system mainly focuses on protecting the textual messages that the users publish on their timelines or walls. The protection of this kind of data is very relevant because it is the main and most immediate source of personal information (Viejo & Sánchez, 2014). Nevertheless, it is worth mentioning that the ideas presented in this paper can also be used to protect other types of contents that are referred by means of textual labels, such as tagged images. We next summarize the main contributions and differentiating features of our privacy-preserving solution:

- Once a configuration step in which the user specifies her privacy requirements is performed, the proposed scheme works in a fully automatic manner. In contrast with the other proposals in the literature, our scheme automatically assesses which textual content to be published by the user must be protected according to its *degree of sensitivity* and the *privacy requirements* of the user. As a result, the user is not required to manually select who can access to each individual resource. To identify the privacy risks that these resources may cause with respect to the privacy requirements of the user, our system relies on an information theoretic assessment of the sensitivity of the data, which has been successfully used in the document sanitization literature to evaluate the disclosure risks of plain text (Sánchez, Batet, & Viejo, 2013a).

- It provides a flexible and intuitive configuration of the trade-off between the privacy and the utility (i.e., readability/informativeness) of the published data according to the *privacy requirements* of the user that owns the data and the readers that access to such data. More specifically, the proposed scheme seamlessly discloses more or less information (thus, achieving less or more privacy) about a certain protected resource according to the credentials of the users that access it (i.e., users with strong access rights can obtain more information than users with weak rights); this contrasts with current solutions, which basically provide full access or no access at all.
- It does not require the collaboration of the social network. In fact, the social network operator is considered an adversary in our scenario. In order to transparently integrate our solution with a non-collaborative social network, we use some of the preliminary ideas presented in (Viejo, et al., 2013). This includes publishing inaccurate data in the social network and hiding the sensitive original data in apparently standard images by means of *steganography* in order to achieve transparency.

The rest of this paper is organized as follows. First, we present and formalize our proposal by detailing how sensitive information is automatically detected, protected and transparently stored and recovered. Then, we discuss the feasibility of the proposed system and illustrate it in two simulated social networks. Finally, we give some concluding remarks and propose some lines of future research.

## 2. The proposed system

First, we briefly describe how our system works using a simple running example. Then, the basic requirements that the targeted social network should fulfill in order to be able to apply our solution are introduced. Finally, we explain our proposal in detail.

### 2.1. The proposed system in a nutshell

Our system is envisaged as an independent software that does not require the collaboration of the social network and that has to be installed in the computers of the users of the social network, both those who publish content and those who read it.

The main goal of the proposed system is to automatically and transparently protect the *textual messages* published by a user in her social network account in a way that each potential reader will be able to only get the amount of information allowed by her credentials, while the social network provider and external readers will be able to only get the most general information. This behavior protects the privacy of the users by preventing non-authorized parties from gathering sensitive information.

For example, let us imagine a user who is using the proposed scheme and is willing to share in her social network the message: "I have a lung cancer". Also, note that the contacts of the user are classified according to three categories: *close-friends*, *friends*, *others*.

First, before allowing this message to be published, the proposed system analyzes it. Let us assume that, as a result of this analysis and taking into account the *privacy requirements* of the user (they will be discussed later), the system considers that the term "lung cancer" is very sensitive and it should only be accessible by *close-friends*. The system also concludes that a generalization of "lung cancer" such as "tumor" would be suitable (because it lowers the amount of disclosed information enough) for *common friends* while any *other entity* (which includes external parties and the social network operator) should only be able to learn the more abstract generalization "disease". As a consequence, the system replaces the term "lung cancer" by "disease" in the publication, uses cryptography to protect the terms "lung cancer" and "tumor" by making them accessible only to the readers with the right credentials and, finally, hides the ciphered data in a certain image (by means of steganography) that is published together with the final message that now only states: "I have a disease".

Now, let us move to the point of view of the reader who is willing to read the protected message published by the user. First, two main situations may apply depending on whether the reader is aware of the privacy-preserving system used to protect the publication or not. In this way, if *the reader is not aware*, she only reads "I have a disease". On the other hand, if *the reader is aware*, she is able to obtain as much information (i.e., "lung cancer" or "tumor") as her credentials and the rules of access in use (which are applied according to the *privacy requirements* of the user) allow her. That is, any non-authorized entity will only be able to read "I have a disease" due to the fact that she will not be able to retrieve any element from the steganographed image by using her credentials/cryptographic material. In contrast, any individual classified as *common friend* will be able to obtain from the steganographed image the textual element "tumor" as a more specific replacement for the protected term. This element combined with the published message results in the final text: "I have a tumor". Finally, *a close friend*, who has complete access to user's information, will be able to obtain the term "lung cancer" from the steganographed image and, hence, she will be able to reconstruct the original text: "I have a lung cancer".

## 2.2. Basic system requirements

The proposed system is designed to protect the sensitive data of a user by substituting them by generalized versions (e.g. *lung cancer -> tumor -> disease*) or even by removing (i.e., redacting) them, while storing the original information in standard images which are published in the user's social network account. Thus, the system requires that the social network provides a place to publish *text messages* or textual tags together with *images*. This feature is commonly supported by most social networks that usually refer to it as *Timeline* or *Wall*.

Moreover, it is assumed that the different readers of a social network account can be classified according to their relationship with the user who owns the account (e.g., friends, relatives, followers, etc.), which should be sorted according to their closeness/trust with her. These relationships are generally supported in all the social networks (B. Carminati, Ferrari, & Perego, 2009) and they are specified by the users themselves when they add a new contact to their list of friends or, in some cases, when the users request to create a new connection with someone (i.e., the case of the followers). It is worth mentioning that the lack of a classification for other entities, such as the social network operator or external entities, represents an implicit classification (i.e., not-authorized entities). In order to perform this operative the social network is required to provide a *list of contacts*.

Examples of already deployed social networks that fulfill all these requirements are Facebook, PatientsLikeMe, Twitter, LinkedIn or Google+, among others.

## 2.3. The proposed system in detail

The enforcement of the proposed system consists on a preliminary step (*Step-0*), that is expected to be run only once and that requires the interaction of the user in order to define her *privacy requirements*, and three main steps (*Step-1*, *Step-2* and *Step-3*, respectively) that are run sequentially and automatically for each new user publication in order to enforce the privacy protection. They are next explained in detail.

### 2.3.1 Step-0: Defining the user's personal privacy requirements

The purpose of this preliminary step is to define the *privacy requirements* of the user, which are meant to be acquired when deploying the system into the user's environment. These requirements are defined as a list of *n* of *privacy levels* that allow classifying the published data according to their degree of sensitivity and are associated to each type of contact in the social network (e.g., *friends*, *followers*, *others*). These will be later used to enforce the protection process by allowing the access to detailed information only to the readers with the appropriate credentials.

Regarding the assessment of sensitivity, we assume that those terms providing/disclosing a large amount of information are likely to be also sensitive. At this respect, several privacy-protection methods for textual data and empirical studies have shown the close relationship between the informativeness of textual terms and their sensitivity (Abril, Navarro-Arribas, & Torra, 2011; Sánchez, et al., 2013a). Specifically, as done in (Sánchez, et al., 2013a), we measure the informativeness of a term *t* according to its *Information Content (IC)*. Numerically, the *IC* of a term *t* is computed as the inverse of its probability of appearance, *p(t)*, in a corpus.

$$IC(t) = -\log_2 p(t) \tag{1}$$

In this manner, general terms such as *disease* are assumed to provide less information than specialized ones such as *lung cancer*, because the former are more likely to be referred in a discourse. Likewise, *lung cancer* is assumed to be more sensitive than *disease*, because of the larger amount of information it discloses. To compute realistic term probabilities, we rely on the largest and most up-to-date electronic repository available: the Web. In fact, the Web is so large and heterogeneous that it is said to be a faithful representation of the information distribution at a social scale (Cilibrasi & Vitányi, 2006). This argument that has been supported by recent works on privacy-protection (Chow, Golle, & Staddon, 2008; Sánchez, et al., 2013a; Sánchez, Batet, & Viejo, 2013b), which considered the Web as a realistic proxy of social knowledge. In order to compute term probabilities from the Web in an efficient manner, several authors (Sánchez, Batet, Valls, & Gibert, 2010; Turney, 2001) have used the *hit count* returned by a Web Search Engine (e.g., Bing, Google) when querying the term *t*. In our approach, term probabilities are computed in this way:

$$IC_{web}(t) = -\log_2 \frac{hits(t)}{N} \tag{2}$$

where *N* is the number of web resources indexed by the web search engine.

By relying on this assessment of sensitivity, the *n privacy levels* are organized in a way that the lowest level $L_0$ is related to the least informative terms that do not require any protection (i.e., they can be accessed by external non-classified readers and also by the social network provider), whereas the highest privacy level $L_{n-1}$ is linked to the most informative ones, which can only be accessed by fully trusted contacts. The access to the sensitive data of a certain user is done by linking each type of contact/relationship of that user (e.g., *common friend*, *close friend*, etc.) to a certain privacy level $L_i$. The main idea behind this access method is that the type of relationship that exists between the owner to the data and each potential reader defines their trust and, thus, the *maximum amount of information* that the latter can obtain from the former's publications. Following the example introduced above, a term such as "lung cancer" (with high-sensitivity) can be classified into the privacy level $i=2$ ($L_2$), while a term such as "tumor" (with medium-sensitivity) can be classified into the level $i=1$ ($L_1$); in the same way, a reader classified as *friend* of the data owner can be allowed to access to any information classified into $L_1$, while an individual classified as *close friend* may get any information belonging to $L_2$.

In our approach, the number of types of relationship between users of the social network defines the number of privacy levels, which are sorted according to their trustfulness. Recalling the previous example, let us assume a certain social network that allows the user to classify each contact as *close-friends* or as *friends*. Note that, there will be also an implicit third type of relationship to which any other entity belongs (i.e., the social network operator, external readers, etc.). As a result, the *privacy requirements* of a user of this specific social network would contain three different privacy levels: $L_0$, $L_1$ and $L_2$, linked to *others*, *friends* and *close-friends*, respectively.

In order to define the specific *privacy requirements* of the user, she is asked to define the amount of information that would be disclosed, at most, for each privacy level/type of relationship. In order to make this process straightforward, and inspired by the usual approach for acquiring requirements in Software Engineering (Hay, 2003), the privacy requirements are automatically setup by answering a set of predefined and intuitive questions about sensitive topics for each type of contact. Specifically, the system presents a set of questions related to different sensitive topics (e.g., religion, race, sexuality, health, etc.) and, for each one, the user has to decide the amount of information that, at most, the readers belonging to each type of relationship/privacy level can obtain. For example, the user can decide that close contacts can get full details about the sensitive diseases of the user (e.g., "HIV"), whereas regular contacts can only get generalized information (e.g., "infection") and other non-tagged contacts can get nothing.

The proposed scheme provides freedom to the system designer to define the topics, questions and answers related to the privacy requirements, so that they can cover a variety of sensitive matters. However, as a recommendation, those should be defined to:

- Capture the notion of privacy that it is described in applicable legal frameworks. For example, the HIPAA specifies that census-related information (e.g., locations) should be protected in medical records (Department of Health and Human Services, 2000); several U.S. federal laws stipulate that conditions such as HIV status, drug or alcohol abuse and mental disorders are sensitive

(Department for a Healthy New York, 2013); finally, the EU Data Protection specifies that any potentially discriminatory information related to religion, political opinions, race or sexual orientation should be protected (The European Parliament and the Council of the EU, 1995).
- Be coherent with the thematic scope of the social network (e.g., social relationships in Facebook, healthcare experiences in PatientsLikeMe, etc.).

Notice that this is the only step of the proposed system that requires the interaction of the user, who has to answer a set of questions. It is worth mentioning that the number of questions is directly linked with the number of topics being considered that, according to the current legislations depicted above, can be summarized in six major topics: Health status (including mental disorders), drugs/alcohol abuse, religion beliefs, political opinions, sexual orientation and census data. As a result, a generic implementation of the proposed method can just ask one general question related to each one of these private topics. Moreover, if the targeted social network focuses on a specific topic, the number of topics to be addressed can be reduced because it may only make sense to ask questions related to the main topic of the social network. On the other hand, the general topics can be extended or redefined so that they are tailored to the scope of the social network (e.g., specific location data in a social network devoted to traveling).

Once the questions have been answered, the informativeness of these answers for each type of contact/privacy level $L_i$ (computed as in eq. (2)) is used as threshold $T_{Li}$ for that level. This threshold will be used to determine, in the next step, the level to which a certain textual term $t$ published by the user (with an informativeness of $IC(t)$) belongs to and, thus, to take the appropriate protection measures according to the credentials of the reader. In other words, the threshold specifies the maximum amount of information disclosure allowed for each type of reader. If different questions about several sensitive topics are performed to the user, the answer with the lowest informativeness for each privacy level (i.e., the one that imposes the strictest protection criterion) will be used as threshold. Note also that a threshold of 0 (i.e., the user does not allow to disclose *any* information) will result in a completely redacted output in which all the terms will be removed, since any term and generalization has an IC>0. On the other hand, an infinite threshold (i.e., the user allows to disclose *all* the information) will produce a non-sanitized output, in which all terms are published in clear because all of them fulfill the threshold.

Summarizing, after this step, the *privacy requirements* of the user are completely determined as a tuple of privacy levels $L_i$, each one with a linked threshold $T_{Li}$ stating the maximum amount of information that is allowed to be disclosed for each type of reader.

## 2.3.2 Step-1: Detecting sensitive terms and obtaining privacy-preserving generalizations

In this step, the system takes as input the message to be published by the user and her *privacy requirements*. Then, it performs several linguistic analyses to extract potentially sensitive terms, whose sensitivity is evaluated by means of their degree of informativeness (eq. (2)) with regard to the thresholds of each privacy level. Finally, the system uses a set of knowledge bases (i.e., WordNet (Fellbaum, 1998) and ODP[3] for general terms and SNOMED-CT (Kuntz & Berkum, 2005) for medical entities) to provide generalizations that can be used as replacements of the terms to be protected (i.e., those that reveal more information that the amount defined by the threshold of the privacy level). The idea is that, by replacing a sensitive term with a generalization (e.g., *lung cancer -> tumor*), we are lowering the amount of information disclosed to a certain type of readers (in order to fulfill the *privacy requirements*) while retaining an amount of its semantics/utility (Sánchez, et al., 2013a).

Due to the fact that the textual messages published in a social network are assumed to lack of a regular structure, the proposed system uses several natural language processing tools to detect sensitive terms. Specifically, because sensitive terms are mostly concepts or instances and these are referred in text by means of noun phrases (NPs) (Sánchez, et al., 2013a), the system focuses on the detection of NPs. NPs are detected by means of several *natural language tools*[4], which perform sentence detection, tokenization (i.e., word detection, including contraction separation), part-of-speech tagging (POS) and syntactic parsing of the input text.

Once NPs are detected and stop words (e.g. prepositions, determinants, etc.) are filtered, each one (which we refer generically as term $t$) is classified according to its level of sensitivity (i.e., the privacy level to which it belongs) and stored in a clear or in a generalized (sanitized) way in a set $S_i$ that contains the information available to readers of level $L_i$. In this manner, readers with credentials linked to the same privacy level of $t$ (or to higher privacy levels) will be able to obtain $t$. For readers with less privileges (i.e., those who are linked to lower privacy levels), the system will provide a generalization of $t$ (i.e., $g(t)$) suitable for the readers' privacy level $L_i$ (i.e., $g_i(t)$ provides low enough information to be classified under $L_i$ because its informativeness is below the threshold $T_{Li}$). Finally, only the terms or generalizations classified in the lowest privacy level ($L_0$) will get published in the social network, thus being available to external entities without credentials and to the network operator. This process is next described more formally.

---

[3] http://www.dmoz.org/ (last accessed: January, 2016)
[4] http://opennlp.apache.org/ (last accessed: January, 2016)

Assuming *privacy requirements* with $n$ levels $\{L_0,…,L_{n-1}\}$ and their corresponding $n$ thresholds $\{T_{L0},…,T_{Ln-1}\}$, for each term $t$ in a certain message $m$ to be published do:

- If $T_{L0} \geq IC(t)$, term $t$ is not sensitive because it provides less (or the same) information than such allowed in the least restrictive threshold $T_{L0}$, which corresponds to external readers with no credentials and with no specific connection with the user:
    - The system does nothing and $t$ will be published as is, so that readers in $L_0$ (or higher) can access to it.
- If $T_{Li+1} \geq IC(t) > T_{Li}$, term $t$ is sensitive for readers in $L_i$ or below:
    - The system obtains the most informative generalization $g_0(t)$ from the knowledge bases in use such that $T_{L0} \geq IC(g_0(t))$.
    - $t$ is replaced by $g_0(t)$ in $m$, so that any entity that belongs to $L_0$ only learns $g_0(t)$ of $t$, or is removed if a suitable $g_0(t)$ is not found.
    - $t$ is included in the sets $\{S_{i+1},…, S_{n-1}\}$ accessible by readers with credentials linked to $L_{i+1}$ or higher.
    - The system obtains the most informative generalization $g_i(t)$ from the knowledge bases in use such that $T_{Li} \geq IC(g_i(t))$.
    - If $g_i(t) \neq g_0(t)$, $g_i(t)$ is included in the set of terms $S_i$ accessible by readers with credentials linked to $L_i$.
    - Repeat the last two steps for privacy levels $L_{i-1}$ (obtain a suitable $g_{i-1}(t)$ and store it, if necessary, into $S_{i-1}$) and below until reaching $L_1$.
- If $IC(t) > T_{Ln-1}$, term $t$ is sensitive for all the contacts of the user:
    - The system obtains the most informative generalization $g_0(t)$ from the knowledge bases in use such that $T_{L0} \geq IC(g_0(t))$.
    - $t$ is replaced by $g_0(t)$ in $m$, so that any external entity that belongs to $L_0$ only learns $g_0(t)$ of $t$, or is removed if a suitable $g_0(t)$ is not found.
    - Obtain and store generalizations for privacy levels $L_{n-1}$ (obtain a suitable $g_{n-1}(t)$ and store it, if necessary, into $S_{n-1}$) and below until reaching $L_1$.

With the above described procedure, $S_i$ will store the terms or the most informative generalizations of the sensitive terms that the *privacy requirements* of the user allow for the contacts linked to each $L_i$. In this manner, the information disclosed to each type of contact will retain the maximum utility allowed by the generalizations available in the knowledge bases in use while fulfilling the *privacy requirements*. Note that there is no $S_0$, since the terms or generalizations associated to that level are directly published in the social network.

Notice that, due to polysemy, some of the terms that should be replaced may correspond to *several* generalizations in the knowledge bases, one for each sense of the term (e.g., *cancer* may refer to a *malignant tumor*, being *disease* and appropriate generalization, or to *astrology*, being *constellation* the suitable one). In order to select the appropriate generalization, a semantic disambiguation process is performed. This process relies on the hypothesis that, to be semantically coherent, the senses of the terms appearing in the same sentence should refer to a common topic, that is, they should be *semantically similar* (Kilgarriff & Rosenzweig, 2000); thus, the context of the terms (i.e., the other terms appearing in the same sentence) can be exploited to perform the disambiguation. Specifically, we select the most appropriate generalization of a term as the one (from the set of possible generalizations) that, in average, is the most semantically similar (according to a state of the art measure (Sánchez, Batet, Isern, & Valls, 2012)) to all the other terms in the same sentence.

### 2.3.3 Step-2: Protecting sensitive data

The sets $\{S_1,...,S_{n-1}\}$ that were generated in *Step-1* are the input of this step. These sets contain the terms or generalizations (i.e., sanitized versions) that are accessible for the users belonging to each privacy level. Note that each term $t$ in $S_i$ has been replaced by $g_0(t)$ in the original message and, since the stored terms and their replacements in the message follow the same order, it is straightforward to link them and, thus, to recover the appropriate text for each privacy level. The goal of this step is to encrypt the terms in $\{S_1,...,S_{n-1}\}$ in a way that each reader of the publication will be able to get only the $S_i$ linked to the privacy level $L_i$ to which she belongs. Encrypted sets will be later stored into images published together with the protected contents via steganography, so that the appropriate information could be retrieved transparently for readers with the appropriate credentials and cryptographic materials.

In order to achieve that, each privacy level $L_i$ (with $i>0$) has a corresponding AES key $K_{Li}$. Each of these keys is used to encrypt, by means of the AES cryptosystem, the corresponding set $S_i$ of terms (original terms or generalizations suitable for the privacy level) that are accessible to the readers with credentials for $L_i$. In this way, $K_{L1}$ is used to encrypt a data structure $DS$ containing all the terms in $S_1$, $K_{L2}$ is used in the same way with $S_2$, and so on. As a result, the system obtains $n$ encrypted elements $C_i$ where $C_i = E_{KLi}(DS_i)$.

Next, the proposed scheme uses a *broadcast encryption technique* to enforce the access to the protected terms according to the credentials of each user. In this way, the keys $\{K_{L1},…, K_{Ln-1}\}$, which are required to decrypt the different sets of terms, are encrypted according to a certain broadcast encryption technique. The resulting element is generally called *Media Key Block (MKB)* in the literature related to broadcast encryption (Naor, Naor, & Lotspiech, 2001). A MKB is essentially a data structure that encrypts a session key (in our case, this is a key $K_{Li}$). Each entity that can try to access this session key owns an *individual key* and uses it to recover the session key from the MKB.

The use of a broadcast encryption technique to enforce the access to the protected contents in the considered scenario is essential due to the following points: (i) social networks are scenarios in which revoking users (i.e., removing contacts of a user) is a common situation, therefore, the proposed method should deal with user revocations in an effective way; (ii) the social network operator is considered to be an adversary, therefore, the proposed system cannot rely on it; and (iii) social network users cannot be assumed to be always on-line, therefore, the proposed method for enforcing access to protected content cannot rely on the existence of a direct dialogue between the owner of the data and the entity trying to get this data. Broadcast encryption methods are specifically suited for dealing with these three points. Basically, they enable the owner of the protected data to grant or revoke the access to one or several readers in an easy and non-interactive way. Even though many broadcast encryption techniques could be used in practice, it is worth mentioning that the well-known *Subset Difference (SD) scheme* (Naor, et al., 2001) was designed to deal efficiently with user revocations and, hence, it can work properly in the considered scenario (Viejo, et al., 2013).

As it has been explained above, the use of broadcast encryption requires the user to share a set of secret keys with her contacts in the social network, who appear in the *list of friends* of the user and are classified according to their relationship with the user (e.g., close friends, common friends, followers, etc.). When an individual is classified (i.e., the user confirms their relationship), the system gives her the required set of secret keys. This can be achieved by means of any independent telematic channel outside the control of the social network. Probably, the simplest approach would be to send/receive those keys by e-mail (email addresses can be directly found in the social network) as suggested in (Viejo, et al., 2013). Nevertheless, the proposed system is not bound to use a specific way of distributing the cryptographic material.

Finally, note that the individuals that do not appear in the list of friends but can still try to get access to the data (i.e., the social network operator, external entities, etc.) do not have any credential and they are classified into the lowest privacy level $L_0$ by default. Hence, they will not be able to get the cryptographic material required to get any sensitive term and they will only learn the most abstract generalizations, $g_0(t)$, which were used as replacements for the protected terms in the published message. On the other hand, the authorized readers (i.e., those belonging to $L_1$ or higher) will be able to use their assigned secret keys to retrieve the corresponding $K_{Li}$ from the MKB and to obtain the original terms or the appropriate generalizations stored in the $S_i$ associated to their credentials.

**2.3.4 Step-3: Hiding the protected data from the social network operator**

The inputs of this step are the set of encrypted elements $\{C_1,...,C_{n-1}\}$ that contain the protected terms and the MKB that will be used by authorized readers to obtain the right cryptographic keys and get the corresponding protected terms. As it has been previously explained, to avoid availability issues or problems inherent to distributed solutions, the proposed scheme stores the protected information in the servers of the social network; however, this data is stored in a way that prevents the social network operator from being aware of its existence. In order to achieve that, the protected data is stored (i.e., hidden) in apparently innocuous images by means of *steganography*; that is, the set of encrypted elements and the linked MKB that correspond to a certain message are steganographed in a single image that is published together with the most protected version of the initial message (which corresponds to $L_0$) in the social network.

The proposed system is not linked to any specific steganographic method. However, it is worth mentioning that the steganographic scheme has to provide good information rate (to be able to store enough information in the image), imperceptibility (in order not to be evident to the social network operator), robustness (in order to withstand the usual compression applied by the social network operator to the uploaded images) and, finally, it has to be oblivious (i.e., the recovery algorithm should not require the original unmarked image).

Regarding the cover images to be used as input to the steganographic method, it is worth mentioning that the graphical contents depicted by images are not relevant (as it has been explained above, their only purpose is to prevent the social network operator from being aware of the encrypted data). Because of this, cover images should be selected according to the maximum quantity of information that they can store as a function of their size, but also according to their capacity to store more or less data because of the nature of their graphical contents. This implies that the proposed system can be deployed with a predefined set of cover images with good and known storage capacities. Then, given a certain message to be protected, the system can select a suitable cover image according to the storage needs raised during the current protection process (i.e., the number of terms to be protected, the privacy levels and the number of contacts). In this manner, we can ensure that the enforcement of the proposed schema is always feasible in practice.

## 3. Empirical study

In this section we discuss the applicability of the proposed system in two types of social networks: a thematic network about healthcare and a generalist microblogging one. These two scenarios have been selected due to their well-differentiated scopes, potential types of relationships between users, and also because of the different limitations they impose on the message length. Specifically, textual publications in microblogging networks such as *Twitter* are at most 140 characters long. Note that, as stated in (Viejo, Sánchez, & Castellà-Roca, 2012), textual messages with a very limited length are more difficult to analyze because their syntactical structures are not as well-formed as for unrestricted messages. The two considered scenarios have been defined to be realistic by mimicking the design and architecture of real social networks.

To show the feasibility of our system, we will simulate a synthetic user $U$ using both social networks. First, the *privacy requirements* of $U$ will be defined for each social network. This implies running the *Step-0* described above, which consists on defining the level of information that will be available to the different contacts of $U$ for a sensitive topic. Then, the behavior of our system will be illustrated with two synthetic but realistic examples of textual publications. After that, the two messages will be processed in the *Step-1,* which is in charge of detecting sensitive terms and of providing privacy-preserving generalizations according to the privacy requirements that have been previously defined. As a result, we will obtain the sanitized output for each type of reader of the publication for each social network. The outputs will be analyzed according to the degree of informativeness they provide. Finally, we will discuss the technical details related to the storage and encryption of sensitive terms into the images, that is, the issues regarding the number of bytes to be stored in each case, the requirements of the images and the information rates of some well-known steganographic methods available in the literature.

### 3.1. Defining the privacy requirements of the user

As described earlier, the system gathers the privacy requirements of the user by means of a set of questions related to different sensitive topics where she has to decide the amount of information that, at most, the readers belonging to each type of relationship can obtain. According to that, the number of different types of contacts supported by the target platform and the number of different sensitive topics to be covered are important variables that directly affect the set of pre-defined questions to be used.

As detailed in (B. Carminati, et al., 2009), most social networks offer, on average, three different types of relationships to their users. Focusing on the two example scenarios, for the healthcare-oriented social network, we considered the following three types of relationships, sorted by their level of trust: "clinician/researcher", "follower" and "registered user". A "clinician/researcher" may be a healthcare professional that uses the data published in the social network for healing/research purposes. A "follower" of a user $U$ is any user who decides to follow $U$ and explicitly states this decision to the social network. Finally, a "registered user" is any user who owns a valid account in the social network and she does not fall into any of the former categories. We assumed that individuals without a valid account cannot access to any information published in this social network; therefore, the social network operator is the only external entity that belongs to the privacy level $L_0$. Regarding the microblogging network, we considered two types of relationships: "follower" and "registered user". However, since we assumed that all the content is publicly available by default, external entities are also capable of reading information published in this social network. Thus, we considered "external entity" as a third type of readers that, due to the fact that they are not related to the user, belong to the privacy level $L_0$ together with the social network operator.

Regarding the number of questions to be asked, as stated earlier, those can be defined according to the elements that are considered to be sensitive in current privacy regulations and the scope of the social network. By focusing in the two considered scenarios, we will illustrate the protection process for one private topic. Given the different scopes of the two social networks in this study, the privacy requirements for the healthcare social network will refer to the *condition* of the user, whereas for the microblogging network, they will refer to the *location* of the user.

Figure 1 shows a sample questionnaire that can be used by the users of the healthcare social network to specify their privacy requirements. To avoid errors or incoherencies in the answers, questions are sorted according to the trust (i.e., privacy level) of the different types of readers, and predefined answers with different degrees of informativeness are given. The answers reflected in this figure correspond to the simulated answers considered for the user $U$.

> With regard to your *condition*, select the maximum information that you are willing to disclose in your messages for each type of contact:
> - Clinician/Researcher: (Level $L_3$)
>   - ☒ she can know everything about your condition
>   - ☐ she can know your condition but not more specific details
>   - ☐ she can know your condition type but not the concrete condition
>   - ☐ she can know that you suffer from a condition, but not the type
>   - ☐ she can know nothing at all
> - Follower: (Level $L_2$)
>   - ☐ she can know everything about your condition
>   - ☒ she can know your condition but not more specific details
>   - ☐ she can know your condition type but not the concrete condition
>   - ☐ she can know that you suffer from a condition, but not the type
>   - ☐ she can know nothing at all
> - Registered user: (Level $L_1$)
>   - ☐ she can know everything about your condition
>   - ☐ she can know your condition but not more specific details
>   - ☒ she can know your condition type but not the concrete condition
>   - ☐ she can know that you suffer from a condition, but not the type
>   - ☐ she can know nothing at all
> - External entities: (Level $L_0$)
>   - ☐ she can know everything about your condition
>   - ☐ she can know your condition but not more specific details
>   - ☐ she can know your condition type but not the concrete condition
>   - ☒ she can know that you suffer from a condition, but not the type
>   - ☐ she can know nothing at all

*Figure 1*. Sample questionnaire for the healthcare oriented social network

Figure 2 shows the questionnaire and the simulated answers for the microblogging social network, which is related to the location of the user.

> Your *location* is sensitive information. Select the maximum information that you are willing to disclose in your tweets for each type of contact:
> - Follower: (Level $L_2$)
>   - ☒ she can know all the details of your location
>   - ☐ she can know the city but not more concrete details
>   - ☐ she can just know your country
>   - ☐ she cannot know your country
> - Registered user: (Level $L_1$)
>   - ☐ she can know all the details of your location
>   - ☒ she can know the city but not more concrete details
>   - ☐ she can just know your country
>   - ☐ she cannot know your country
> - External users: (Level $L_0$)
>   - ☐ she can know all the details of your location
>   - ☐ she can know the city but not more concrete details
>   - ☒ she can just know your country
>   - ☐ she cannot know your country

*Figure 2.* Sample questionnaire for the microblogging social network

Note that these sample questions (and answers) are just examples that are used to illustrate the behavior of the proposed system. Designing better questions or providing a better layout is out of the scope of this paper. Note also that there could be as many possible answers as desired (or even allow a free textual input stating a concept with a specific degree of informativeness), and that the same level of information disclosure could be associated to different contacts/privacy levels; in this latter case, the thresholds of the different levels, and thus, the resulting sanitization, would be the same.

According to the answers of the user, the privacy requirements are set by computing the threshold $T_{Li}$ for each privacy level $L_i$. As it has been previously explained, thresholds are computed by using the IC of the answers provided by the user to the set of predefined questions.

In the questionnaire for the healthcare scenario, thresholds can be directly defined according to the condition that the user $U$ defines in her profile in the social network and the answers to the questions with regard to that condition. Let us assume that $U$ claims to have "HIV" (which is sensitive according to (Department for a Healthy New York, 2013)). Therefore, according to her answers, $L_3$ readers can know everything, thus $T_{L3}$ is infinite; $L_2$ readers can know at most her sensitive condition, thus, $T_{L2}$ is computed according to the informativeness of HIV[5]: $T_{L2} = IC(\text{"HIV"}) = -\log_2(15E6/17E9) = 10.14$; $L_1$ readers can only know about the condition type, which, according to the generalizations of HIV in the knowledge bases in use would correspond to "Infection", thus, $T_{L1} = IC(\text{"Infection"}) = - \log_2(16.5E6/17E9) = 10.0$; finally, any external entity, which in this case would only be the social network operator that corresponds to $L_0$, will not even know the type, that is $T_{L0} = IC(\text{"Condition"}) = - \log_2(17.5E6/17E9) = 9.92$. The last answer in the questionnaire, which does not allow disclosing any information at all, would correspond to $T_{Li} = 0$, which would produce a completely redacted output in which all terms would be removed.

---
[5] IC is computed using eq. (2) and Bing as the Web Search Engine to retrieve the number of hits. The number of web resources indexed by Bing (in the denominator) is set to 17 billions, as estimated in http://worldwidewebsize.com (accessed: July 2015)

Regarding the questions related to the microblogging scenario, we can use the location specified by the user in her profile. Let us assume that $U$ is located in "Barcelona" (Spain). Thus, according to her answers, $L_2$ readers can know all the details, thus $T_{L2} = \infty$; whereas $L_1$ readers can only know the city, thus $T_{L1} = IC(\text{"Barcelona"}) = - \log_2(202E6/17E9) = 6.3$; and external readers and the social network provider would only have access to the country, thus, $T_{L0} = IC(\text{"Spain"}) = - \log_2(233E6/17E9) = 6.18$. The latter answer of the questionnaire, in which the country should not be revealed, would correspond to a threshold $T_{Li} > IC(\text{"Spain"})$.

### 3.2. Sanitizing user messages

When the user tries to publish a new message in the social network, the proposed system intercepts this action and performs a linguistic analysis to extract potentially sensitive terms according to their degree of informativeness with respect to the defined privacy requirements. For each detected term, the system retrieves generalizations that are suitable for the different privacy levels.

To illustrate this process, let us assume that $U$ posts in the healthcare social network the synthetic message about her condition that is shown in the first row of Table 1. Likewise, the first row of Table 2 shows a synthetic message related to Barcelona that is intended for the microblogging network. Within the running examples, we represent generalizations with all capital letters so that the reader can easily distinguish them from the text that is left untouched. Notice, however, that in a real setting these generalizations will be published with a regular capitalization so that they do not increase the attention of external entities over the protected terms.

Table 1

*Input, syntactically analyzed and protected messages (for different privacy levels) in the healthcare social network and their percentage of information preservation.*

| Message type | Content | Info |
|---|---|---|
| Input | I've got HIV in 2008. In June 2008 I've got a pharyngitis that stayed for 10 days. After that, I had several ulcers in the mouth. Suspecting an infection, I went to the hospital and the physician asked for an HIV testing that was positive. My immune system was very weak, I got pneumonia 3 times. | - |
| Analyzed | I've got [HIV] in 2008. In [June 2008] I've got a [pharyngitis] that stayed for [10 days]. After that, I had [several ulcers] in the [mouth]. Suspecting an [infection], I went to the [hospital] and the [physician] asked for an [HIV testing] that was positive. My [immune system] was very weak, I got [pneumonia] [3 times]. | - |
| $L_3$ protected ($T_{L3}=\infty$) | I've got HIV in 2008. In June 2008 I've got a pharyngitis that stayed for 10 days. After that, I had several ulcers in the mouth. Suspecting an infection, I went to the hospital and the physician asked for an HIV testing that was positive. My immune system was very weak, I got pneumonia 3 times. | 100% |
| $L_2$ protected ($T_{L2}=IC(\text{"HIV"})$) | I've got HIV in 2008. In JUNE I've got a DISEASE that stayed for 10 days. After that, I had PATHOLOGY in the mouth. Suspecting an infection, I went to the hospital and the physician asked for an TESTING that was positive. My SYSTEM was very weak, I got DISEASE 3 times. | 77.9% |
| $L_1$ protected ($T_{L1}=IC(\text{"Infection"})$) | I've got INFECTION in 2008. In JUNE I've got a DISEASE that stayed for DAYS. After that, I had PATHOLOGY in the mouth. Suspecting an infection, I went to the hospital and the DOCTOR asked for an TESTING that was positive. My SYSTEM was very weak, I got DISEASE TIMES. | 72.2% |
| $L_0$ published ($T_{L0}=IC(\text{"Condition"})$) | I've got CONDITION in 2008. In JUNE I've got a DISEASE that stayed for DAYS. After that, I had CONDITION in the STRUCTURE. Suspecting an CONDITION, I went to the hospital and the DOCTOR asked for an TESTING that was positive. My SYSTEM was very weak, I got DISEASE TIMES. | 64.1% |

Table 2.

*Input, syntactically analyzed and protected messages (for different privacy levels) in the microblogging network and their percentage of information preservation.*

| Message type | Content | Info |
|---|---|---|
| Original | I will be visiting Barcelona on June 16th to assist to the Accenture Digital Conference with Accenture Spain and key stakeholders. | - |
| Analyzed | I will be visiting [Barcelona] on [June 16th] to assist to the [Accenture Digital Conference] with [Accenture Spain] and [key stakeholders]. | - |
| $L_2$ protected ($T_{L2}=\infty$) | I will be visiting Barcelona on June 16th to assist to the Accenture Digital Conference held by Accenture Spain and key stakeholders. | 100% |
| $L_1$ protected ($T_{L1}=IC(\text{“Barcelona”})$) | I will be visiting Barcelona on JUNE to assist to the CONFERENCE with SPAIN and PERSON. | 34.1% |
| $L_0$ published ($T_{L0}=IC(\text{“Spain”})$) | I will be visiting CITY on JUNE to assist to the GROUP with SPAIN and PERSON. | 27.5% |

First, the two sample messages are syntactically analyzed by means of natural language processing tools in order to detect all the terms (noun phrases without stop words) that may refer to sensitive concepts. Those are shown within brackets [] in the second row of Tables 1 and 2. Then, according to the *privacy requirements* of $U$ in each social network and the IC of the detected terms, the latter are classified into each privacy level and protected (i.e., generalized or removed) in coherence with the considered thresholds. All the protected terms that should be stored in the attached image for each privacy level ($S_i$, where $i > 0$) are listed in Tables 3 and 4. These tables also show the length in bytes of the data to be stored: 1 byte for each character of each term plus and additional byte per term that is used as delimiter (\0). By reconstructing the message from the one published in the social network (which corresponds to $L_0$) with the appropriate credentials, the texts shown in Tables 1 and 2 are obtained, which are linked to each type of reader with the related credentials (i.e., $L_i$, where $i > 0$). In order to illustrate the amount of information disclosed for each privacy level, we measured and included in the last columns of Tables 1 and 2, the percentage of information that is preserved for each protected message *m'* with respect to the original message *m*; this is computed as the ratio between the sum of the IC of each protected and original term in *m'* and *m*, respectively (Sánchez, et al., 2013a):

$$Information\_preservation(m') = \frac{\sum_{\forall t'_j \in m'} IC(t'_j)}{\sum_{\forall t_j \in m} IC(t_j)} \times 100 \qquad (3)$$

By looking at Table 1 and recalling the privacy requirements of $U$ in the healthcare social network, we have that $L_3$ readers (*clinicians/researchers*) would have access to all the details (i.e., $T_{L3} = \infty$); thus, since $IC(t_j) < T_{L3}$ for all terms $t_j$, the whole message will be accessible for those readers, as shown in the third row in Table 1. Consequently, the last column of Table 3 shows that all the sensitive terms will be stored in $S_3$ in order to reconstruct the message for readers in $L_3$. On the other hand, $L_2$ readers would be able to learn that the user has HIV (as shown in the fourth row in Table 1) and thus, $T_{L2} = IC(\text{"HIV"}) = -\log_2(15E6/17E9) = 10.14$; however, more specific terms such as *pharyngitis* or *pneumonia*, whose $IC(t_j) > T_{L2}$ (i.e., $IC(\text{"pharyngitis"}) = 13.42$ and $IC(\text{"pneumonia"}) = 10.49$) will be protected. To do so, the system retrieves their generalizations from the knowledge bases in use, and selects the most informative generalizations $g_2(t_j)$ that fulfill the considered threshold. In this case, the generalization is *Disease* ($IC(\text{"Disease"}) = 9.76$) in both cases. This term is thus stored in $S_2$ (shown in the third column in Table 3) in order to reconstruct the message corresponding to readers in $L_2$ (shown in the fourth row in Table 1). The same protection and storage process is done for $L_1$ (in which, as shown in the fifth row in Table 1, *HIV* is replaced by *Infection* in coherence with $T_{L1}$); terms and generalizations for that level are shown in the second column in Table 3. Finally, the message corresponding to $L_0$, which is the one providing the least amount of information, is published in the social network (as shown in the first column of Table 3). Note that, for efficiency, $S_i$ sets only store those generalizations that are not already published in the social network. Moreover, Table 1 shows how the percentage of information preservation lowers proportionally with the privacy level/threshold. This is coherent with the fact that the least trusted readers should access to the least detailed data.

Tables 2 and 4 show the results for the microblogging scenario. In this case, most of the terms in the message are related to the location of the user, and are more specific than the thresholds (*Barcelona* and *Spain*). For $L_1$, most of the terms are protected except *Barcelona*, since $T_{L1}$ allows disclosing the city name. For $L_0$, even *Barcelona* is replaced by *City*, because $T_{L0}$ only allows disclosing the country (*Spain*), which is more general than the city name. Compared with the healthcare scenario, in this case, the percentage of preserved information is much lower because of the more general thresholds. This causes that: (i) most of the terms are protected because they are more informative than the thresholds; and (ii) they are replaced by more general (i.e., less informative) generalizations.

Table 3.

*Terms and generalizations published in the social network and stored in the $S_i$ corresponding to each privacy level $L_i$ (with a \0 delimiter); a single \0 also states that the generalization or term for $S_i$ is the same as such already published. The last row shows the size (in bytes) of each $S_i$ to be stored in the attached images.*

| Published | $S_1$ | $S_2$ | $S_3$ |
|---|---|---|---|
| CONDITION | INFECTION\0 | HIV\0 | HIV\0 |
| JUNE | \0 | \0 | June 2008\0 |
| DISEASE | \0 | \0 | pharyngitis\0 |
| DAYS | \0 | 10 days\0 | 10 days\0 |
| CONDITION | PATHOLOGY\0 | PATHOLOGY\0 | several ulcers\0 |
| STRUCTURE | mouth\0 | mouth\0 | mouth\0 |
| CONDITION | infection\0 | infection\0 | infection\0 |
| hospital | \0 | \0 | \0 |
| DOCTOR | \0 | physician\0 | physician\0 |
| TESTING | \0 | \0 | HIV testing\0 |
| SYSTEM | \0 | \0 | immune system\0 |
| DISEASE | \0 | \0 | pneumonia\0 |
| TIMES | \0 | 3 times\0 | 3 times\0 |
| *Bytes to store*: | 45 | 62 | 120 |

Table 4.

*Terms and generalizations published in the social network and stored in the $S_i$ corresponding to each privacy level $L_i$ (with a \0 delimiter); a single \0 also states that the generalization or term for $S_i$ is the same as such already published. The last row shows the size (in bytes) of each $S_i$ to be stored in the attached images.*

| Published | $S_1$ | $S_2$ |
|---|---|---|
| CITY | Barcelona\0 | Barcelona\0 |
| JUNE | \0 | June 16th\0 |
| GROUP | CONFERENCE\0 | Accenture Digital Conference\0 |
| SPAIN | \0 | Accenture Spain\0 |
| PERSON | \0 | key stakeholders\0 |
| *Bytes to store*: | 24 | 82 |

## 3.3. Storing and hiding protected elements

This last subsection analyzes whether it is feasible to store all the protected elements in a steganographed image and upload it to the corresponding social network. For the sake of concreteness, we assume that each message is published together with an image containing all the protected terms. Note that publishing messages together with images is a common feature in social networks.

The size in bytes of the information that has to be hidden in an image directly depends on two elements: (i) the size of the media key blocks (MKBs) needed to enforce the access to the sets $S_i$ that contain the sensitive terms; and (ii) the total size of all the sensitive terms that are stored in the sets $S_i$.

Regarding the first element, the size of a MBK is strongly related to the broadcast encryption technique in use and also to the number of revoked contacts that the system has to deal with. In order to provide a realistic estimation, in this study, we consider the *Subset Difference (SD)* scheme (Naor, et al., 2001). This is a very well-known method that is especially suited to deal with user revocations. Note that this characteristic is very relevant due to the fact that our proposal applies user revocations to prevent unauthorized readers from retrieving sensitive terms; therefore, it is usual to have a significant number of revoked users at each privacy level. According to (Naor, et al., 2001), a MKB based on this method contains *1.38R* ciphertexts on average, being *R* the number of revoked users. Each ciphertext is the output of a block cipher and it encrypts the corresponding session key that requires protection (i.e., $K_{Li}$ in our case). Note that, within the context of social networks, a revoked user is a reader that is unauthorized to read the sensitive terms belonging to a specific privacy level. According to the study of the anatomy of the Facebook social graph presented in (Ugander, Karrer, Backstrom, & Marlow, 2011), the average friend count in this social network is 190. For the sake of simplicity, let us assume that this value can be extrapolated to other social networks and that these contacts are equally divided between the privacy levels of the social network. Now, let us consider a social network with *n* privacy levels *{$L_1$,..., $L_n$}* that require potential readers to have appropriate credentials. As a result, *n* different MKBs are needed and each one has to provide access to approximately 190/*n* contacts; therefore, each MKB has to deal with about 190 - 190/*n* revoked ones. It is worth mentioning that this is the initial and simplified configuration required to guarantee that only readers with the right credentials can access the terms stored in the corresponding set $S_i$. The owner of the data can decide to revoke more or less contacts at her will. According to the values provided above, a MKB for 190 - 190/*n* revoked contacts contains about 1.38 × (190 - 190/*n*) ciphertexts. Taking into account that each ciphertext is the output of an AES-128 encryption and that the length of a key $K_{Li}$ allows it to be encrypted in a single block, the result is that each required

MKBs has a size in bits of: $128 \times 1.38 \times (190 - 190/n)$. Therefore, for $n$ required MKBs, the total size in bits will be $n \times 128 \times 1.38 \times (190 - 190/n)$. This equation shows that the total space required by the MKBs grows linearly with the number of privacy levels required in the application environment (e.g., $n=5$ requires 16.38KB, while $n=10$ requires 36.87KB). Even though this point may represent a scalability issue, it is worth recalling that, according to (B. Carminati, et al., 2009), the most widely used social networks offer, on average, three different types of relationships to their users, which implies that, in practice, the number of privacy levels (i.e., the value of $n$) to deal with is expected to be small.

Focusing on the scenarios that we consider, in the healthcare social network there are three different privacy levels *{$L_1$, $L_2$, $L_3$}* and, therefore, three different MKBs are needed. As a result, 8.2KB (i.e., $3 \times 128$ bits $\times 1.38 \times 127 = 8.2$KB) is the total space required to store all the MKBs together. On the other hand, the microblogging social network only has two privacy levels *{$L_1$, $L_2$}* and, hence, this case requires a total size of 4.09KB.

Regarding the total size of all the protected terms that are stored in the sets $S_i$, Tables 3 and 4 show that the three sets used in the healthcare scenario and the two sets used in the microblogging scenario require 227 bytes and 106 bytes, respectively. Clearly, these sizes are negligible in comparison with the requirements of the MKBs. Moreover, both values can be assumed to be average sizes for common messages published in social networks. Even though larger publications could be possible, this situation is hardly common in these environments; note also that, in *Twitter,* for example*,* messages are at most 140 characters long. Therefore, the sizes of these elements will not be taken into account in the rest of this study.

As it has been explained above, about $n \times 128 \times 1.38 \times (190 - 190/n)$ bits of data have to be hidden in an image by means of steganography. The proposed system is not linked to any specific steganographic method but, in order to give empirical results, we consider the solution presented in (Viejo, et al., 2013). This method divides the cover image in cells of *AxA* pixels, being *A* an integer from 1 to 8 (i.e., 1 for low robustness, 8 for high robustness). Then, each cell is analyzed. If a cell is homogeneous (all pixels are similar), one bit of information is embedded; if not, it is discarded. Finally, for each selected cell, if a "1" has to be embedded, the least significant bits of each pixel are replaced with a certain fixed pattern *a*; otherwise, if a "0" has to be embedded, these bits are replaced with a certain fixed pattern *b*. Additionally, Reed-Solomon correcting codes (Reed & Solomon, 1960) are used to improve the robustness. Specifically, 20% of the selected cells are used to store correcting codes. As a result, a standard RGB cover image of 1024x1024 pixels (which can be uploaded to most social networks) is capable of storing

102.4KB of information using cells of 1x1 (i.e., one bit of information is stored in a single pixel), 25.6KB using cells of 2x2 pixels, 6.4KB using cells of 4x4 pixels and, finally, 1.6KB using cells of 8x8 pixels.

According to these values, the proposed system is capable of working with cells of 8x8 (high level of robustness) if only one privacy level $\{L_1\}$ is considered (i.e., $n = 1$); it can work with cells of 4x4 (that is robust enough to support high compression methods such as the ones applied by Facebook) if only two privacy levels $\{L_1, L_2\}$ are considered; it can work with cells of 2x2 if up to seven privacy levels are addressed (i.e., $n \leq 7$); and, finally, it is capable of working with cells of 1x1 if up to 26 privacy levels are considered.

Focusing again on the considered examples, all the data required in the healthcare scenario (i.e., 8.2KB) can be stored without problems in the considered cover image using cells of up to 2x2 pixels. It is worth mentioning that this social network does not apply any modification to the uploaded images, therefore, even using cells of 1x1 would be good enough to work properly in this specific environment. Regarding the 4.09KB of data to be hidden microblogging scenario, this amount of information can be stored in the cover image using cells of up to 4x4 pixels which, as it has been previously mentioned, provide a good level of robustness.

## 4. Conclusions and future work

This paper has addressed the problem of preserving the privacy of the users in already established and widely-used social networks. First, two significant aspects that have not received enough attention in the literature have been identified: (i) there is a lack of effective mechanisms that enable users to quantify how dangerous a certain tweet/timeline post is from the privacy point of view; and (ii) the privacy settings usually implemented in social networks offer a very rigid access control. Then, a new privacy-preserving scheme specifically designed to enable a transparent protection of user's publications in front of the different types of readers has been presented.

The new system is envisaged as a module to be installed in the computers of the users (both data owners and readers), and it does not require the collaboration of the social network. In fact, the proposed scheme is completely independent of the social network in use and it only requires from the network a basic set of features such as a place to publish text messages (e.g., Timeline or Wall) and a place to publish images (e.g., Photo Album). As a result, users would interact with their social network through the installed application that would be in charge of protecting textual content to be published from unauthorized readers and also of providing the allowed content for authorized readers.

Our system enables users to configure their personal privacy requirements and apply them to obtain a flexible trade-off between the privacy and the utility of the accessible content. Thus, it is capable of seamlessly disclose more or less information about a certain protected resource according to the credentials of the users who access it.

Moreover, the system is fully automatic once it has been configured according to user's privacy requirements: it automatically assesses which content to be published by the user must be protected according to its degree of sensitivity and her privacy requirements. It is worth mentioning that users do not require any kind of technical knowledge to configure their privacy requirements, since such requirements can be intuitively defined by answering questions, and the whole protection process is transparent. Moreover, the privacy requirements can be made coherent with current legislations on data privacy.

The suitability and applicability of the proposed system and its several modules have been discussed in general and illustrated in two realistically simulated social networks with different scopes and features. In these scenarios, concrete solutions for storing and hiding the elements to be protected have been proposed and evaluated.

Regarding future work, several limitations of the proposed scheme can be tackled in order to make it more practical and accurate:

- A main shortcoming is that the system evaluates the sensitivity of terms independently. As noted in works such as (Anandan & Clifton, 2011; Sánchez, Batet, & Viejo, 2014b), semantic correlations between textual terms can cause additional disclosure risks (e.g., treatments or drugs closely related to a sensitive disease can disclose the latter). Additional semantic analyses are required to deal with this issue (Sánchez & Batet, 2016; Sánchez, et al., 2014b).
- Another issue that may negatively affect the assessment of the sensitivity of terms is language ambiguity. Because web queries submitted to the search engine involve words rather than concepts, language ambiguity (e.g., polysemy) may result in inaccurate probabilities. To palliate this problem, web queries can be contextualized by concatenating potentially ambiguous words with their generalizations (Sánchez, et al., 2010). This ensures that the number of term appearances gathered from the search engine is not ambiguous, thus improving the accuracy of the sensitivity assessment (Sánchez, Batet, & Viejo, 2014a). Even though, the omission of a number of valid appearances may produce data sparseness for very specific terms. Because of this last issue, the actual influence of this disambiguation mechanism in the results should be carefully studied. As an alternative to corpora-based probability calculation that avoids the problem of language ambiguity, very recently, some authors have proposed models to measure the informativeness of terms according to their degree of specificity in an ontology (Batet, Harispe, Ranwez, Sánchez, & Ranwez, 2014; Sánchez & Batet, 2012). Because we use ontologies to retrieve generalizations, we also plan to incorporate and evaluate these mechanisms in our system.

- In order to improve the practical applicability of the proposal, we also plan to adapt and localize the system to specific social networks by studying their scopes and the applicable legal frameworks, and by engineering generic sets of suitable questions to gather the privacy requirements.

**Acknowledgements**

The authors are solely responsible for the views expressed in this paper, which do not necessarily reflect the position of UNESCO nor commit that organization. This work was partly supported by the European Commission under H2020 project "CLARUS"; by the Spanish Government through projects ICWT TIN2012-32757 and "SmartGlacis" TIN2014-57364-C2-R; and by the Government of Catalonia under grant 2014 SGR 537.